\documentclass[aps,prd,twocolumn,superscriptaddress]{revtex4-2}

\usepackage{graphicx}
\usepackage{amsmath}
\usepackage{amssymb}
\usepackage{siunitx}
\usepackage{hyperref}
\hypersetup{
	colorlinks = true,
	linkcolor = blue,
	citecolor = blue,
	urlcolor  = blue,
}
\usepackage{color}
\usepackage[normalem]{ulem}

\begin{document}

\title{WKB energy levels in gapped graphene under crossed electromagnetic fields}

\author{I.O.~\surname{Nimyi}}
\affiliation{Bogolyubov Institute for Theoretical Physics, National Academy of Science of Ukraine, 14-b Metrologichna Street, Kyiv, 03143, Ukraine}

\author{S.G.~\surname{Sharapov}}
\affiliation{Bogolyubov Institute for Theoretical Physics, National Academy of Science of Ukraine, 14-b Metrologichna Street, Kyiv, 03143, Ukraine}

\author{V.P.~\surname{Gusynin}}
\affiliation{Bogolyubov Institute for Theoretical Physics, National Academy of Science of Ukraine, 14-b Metrologichna Street, Kyiv, 03143, Ukraine}
\date{\today }

\begin{abstract}
We consider  a single layer of graphene subjected to a magnetic field $H$ applied perpendicular to the layer
and an in-plane constant radial electric field $E$. The Dirac equation for this configuration does not
admit analytical solutions in terms of known special functions. Using the WKB approximation, we
demonstrate that for gapped graphene the Bohr-Sommerfeld quantization condition for eigenenergies includes an additional
valley-dependent geometrical phase. When this term is accounted for, the WKB approximation exhibits good agreement with results from
the exact diagonalization method except to the lowest Landau level.

\end{abstract}



\maketitle

\section{Introduction}

We are pleased to dedicate this work to the 80th anniversary of academician Vadym Loktev.
He is widely recognized for his significant contributions to condensed matter physics, particularly in the field of graphene physics.
We express our gratitude to him for the collaboration, insightful discussions, unwavering support, and encouragement.
We wish him many years of good health and continued success in his work in theoretical physics.

\begin{figure}[!h]
\includegraphics[width=.4\textwidth]{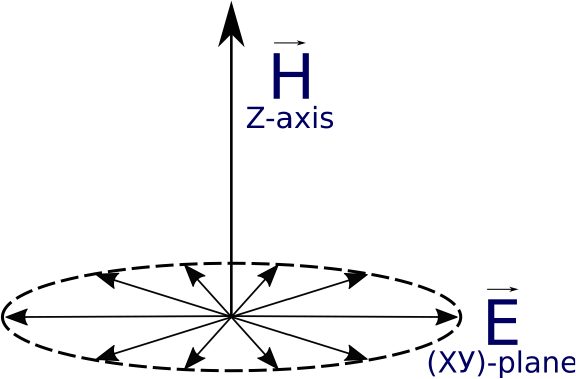}
\caption{A schematic illustration of the electric and magnetic field configuration:
the radial electric field $E$ lies within the plane of the graphene sheet,
while the uniform magnetic field $H$ is oriented perpendicular to this plane.}
\label{fig:1}
\end{figure}

A model for a single layer of graphene
with a magnetic field $H$ applied perpendicular to the layer
and an in-plane constant radial electric field was proposed for investigation
in Ref.~\cite{Nimyi2022PRB} (see Fig.~\ref{fig:1}).
In practice, an approximately constant radial electric field can be generated within a cylindrical capacitor,
where the electric potential is given by
\begin{equation}
V(r) = V_0 \ln \frac{r}{a} \approx V_0 \left(\frac{r}{a}-1 \right),
\qquad b \leq r \leq a,
\end{equation}
with $a$ and $b$ representing the external and internal radii, respectively.
We suggested that this field configuration could also enable the experimental observation of the Landau-level collapse phenomenon.
This effect has previously been reported in Refs.~\cite{Singh2009PRB,Gu2011PRL} using a rectangular geometry with a unidirectional
in-plane electric field.

It turned out that, unlike the case of rectangular geometry with a unidirectional in-plane electric field,
which is exactly solvable \cite{Lukose2007PRL, Peres2007JPCM}, this seemingly straightforward generalization for two-dimensional
(2D) Dirac fermions in a constant magnetic field combined with a constant radial electric field renders
the problem analytically unsolvable.
The situation is analogous to that of 2D Dirac fermions in a constant magnetic field together with a Coulomb potential (see, e.g., Refs.~\cite{Khalilov2000PRA,Gamayun2011PRB,Zhang2012PRB,Sun2014PRB,Moldovan2017-2D,Gorbar2018FNT}) or a parabolic potential $V(r) \sim r^2$
\cite{Rodriguez-Nieva2016PRB}, both of which lack analytical solutions in terms of known special functions.

To address this problem, we employed the semiclassical Wentzel-Kramers-Brillouin
(WKB) method in Ref.~\cite{Nimyi2022PRB}. The WKB approximation for the 2D massless Dirac fermions was discussed
earlier in Ref.~\cite{Zhang2012PRB} in a constant magnetic field.
Specifically, we utilized the Bohr-Sommerfeld (BS) quantization condition for eigenenergies derived in Ref.~\cite{Zhang2012PRB}.
From this, we obtained a transcendental equation for the spectrum of gapless (massless) Dirac fermions in graphene,
expressed in terms of complete elliptic integrals.
These results were compared with calculations obtained via exact diagonalization and shooting methods.
It was found that the WKB solutions closely match the numerical results for nearly all quantum numbers.

The case of gapped (massive) Dirac fermions was studied exclusively through numerical methods,
revealing that the resulting spectra are valley-dependent. However, at the level of the WKB approximation used,
valley distinctions could not be resolved.
The aim of this work is to extend the BS quantization condition for eigenenergies,
making it applicable to the case of 2D massive Dirac fermions.

There are two primary methods for deriving the BS quantization condition:
method which relies on the connection of WKB wave functions between classically allowed and forbidden regions
and the approach based on ensuring the single-valuedness of the WKB wave function.
For example, Zwaan's method  (see the review \cite{Berry1972RPP} and references therein, as well as the textbooks
\cite{Kemble2005book,Landau_vol3book})
is based on analytic continuation of the asymptotic solutions
avoiding turning points  by going around them in the complex plane.
Its application to Dirac fermions is discussed in Ref.~\cite{Zhang2012PRB}.
The second method requires exact solutions near turning points, such as the Airy function, which enables the connection
of regions on both sides of the turning points along the real axis (see the review \cite{Berry1972RPP} and the textbook \cite{Davydov.book}).

However, a more elegant method was proposed by Wentzel \cite{Wentzel1926ZP} and Dunham \cite{Dunham1932PhysRev}
for deriving an eigenvalue quantization condition (see also Refs.~\cite{Froman1977JMP,Parisi1979book}).
This approach utilizes the single-valuedness of the wave function
in the classically allowed region, accurate to all orders in the Planck constant $\hbar$.
Its application to the massless Dirac fermions is discussed in Ref.~\cite{Zhang2012PRB}
(see also Ref.~\cite{Kormanyos2008PRB}) and we chose it for the present work.

The paper is organized as follows.
In Sec.~\ref{sec:model},
we introduce a model for a single layer of graphene subjected to a perpendicular magnetic field $H$ and a constant in-plane radial electric
field $E$. By leveraging the symmetry of the problem, the model is simplified to a system of radial equations.
This system is considered using the WKB method in Sec.~\ref{sec:WKB}
with the technical details are provided in  Appendix~\ref{sec:Appendix-WKB-wave}.
The derivation of the generalized BS quantization condition is presented in Sec.~\ref{sec:Bohr}.
In Sec.~\ref{sec:results}, we obtain and discuss the energy spectra
obtained in the WKB  approximation and compare them with the
results of numerical computations performed using the exact diagonalization method.
In the Conclusion  (Sec.~\ref{sec:conclusion}), we summarize the obtained results.

\section{Model}
\label{sec:model}

We consider the same model Hamiltonian as in Refs.~\cite{Nimyi2022PRB, Herasymchuk2024PRB}
which splits into a pair of two independent
Dirac equations $H_\eta \Psi_\eta( \mathbf{r}) = \mathcal{E} \Psi_\eta( \mathbf{r})$ for each $\mathbf{K}_{\eta}$ point with $\eta = \pm$
with the Hamiltonian density
\begin{equation}
\label{Dirac-eq-2*2}
H_\eta =
- i  \hbar v_F \eta (\sigma_1 D_x + \sigma_2 D_y) + \eta \Delta \sigma_3 +
V(\mathbf{r}) .
\end{equation}
Here, the Pauli matrices $\sigma_i$
act on the ($A,B$) indices of the two-component spinors
$\Psi_+^T = \left( \psi_{AK_+}, \psi_{BK_+} \right)$ and $\Psi_-^T  =
\left( \psi_{BK_-}, \psi_{AK_-} \right)$.
In this work we focus on the case of the massive Dirac fermions with a mass (gap) $\Delta$
which is related to the carrier density imbalance between the $A$ and $B$ sublattices
and does not break the time-reversal symmetry.

The orbital effect of a perpendicular magnetic field, $\mathbf{H} = \nabla \times \mathbf{A}$, is accounted
for through the covariant spatial derivative
$D_j=\partial_j+(ie/\hbar c)A_j$, where $j=x,y$ and $-e<0$.
The potential $V(\mathbf{r})$ corresponds to the static electric field, $e \mathbf{E} =\nabla V(\mathbf{r}) $.
The Zeeman interaction is neglected in this paper due to its negligible effect for moderate magnetic field strengths
(see, e.g., Ref.~\cite{Gusynin2007IJMPB}).

We consider a configuration of crossed magnetic and electric fields, where the magnetic field is applied perpendicular
to the infinite plane of graphene along the positive $z$-axis. The corresponding vector potential is expressed in
the symmetric gauge as $(A_x,A_y)=(H/2)(-y,x)$  The electric field is radial and in-plane, with a potential given
by $V(r)=eEr$  (see Fig.~\ref{fig:1}).

It is clear that the solution at the $\mathbf{K}_-$ point is obtained from the solution
at the $\mathbf{K}_+$ point by changing $\eta \to -\eta$
(which corresponds to the replacement $\Delta \to - \Delta$ in the final equations)
and exchanging the spinor components $\psi_A  \leftrightarrow \psi_B$, so in what follows we only consider
the $\mathbf{K}_+$ point and omit the valley index.

Since the system has rotational symmetry,
it is natural to consider the problem in polar coordinates, where
\begin{equation}
\label{derivative-angle}
i D_{x} \pm D_{y}=e^{\mp i\phi}\left(i\frac{\partial}{\partial r}
\pm \frac{1}{r}\frac{\partial}{\partial\phi} \pm \frac{ie H r}{2\hbar c}\right).
\end{equation}
Accordingly, the total angular momentum
$J_z$ is conserved and we can represent $\Psi(\mathbf{r})$ in terms of the eigenfunctions of
$J_z=L_z+\sigma_z/2=-i\partial/\partial\phi+\sigma_z/2$ as follows:
\begin{equation}
\label{angular-spinor}
\Psi(\mathbf{r})= \frac{1}{\sqrt{r}} \left[\begin{array}{c}e^{i(j-1/2)\phi}F(r)\\ i e^{i(j+1/2)\phi}g(r)\end{array}\right],
\end{equation}
where $j=\pm 1/2,\pm 3/2,\ldots$ is the total angular momentum quantum number.
Then for the spinor
$\Phi(r)=\left( F(r), G(r)\right)^T$,
we obtain the following system of equations written in a matrix form:
\begin{equation}
\label{WKB-Dirac}
\hbar \Phi'(r)= M \Phi(r),
\end{equation}
where the matrix
\begin{equation}
\label{matrix-D}
\begin{split}
M & = \left[\begin{array}{cc}\frac{J}{r}+
\frac{e H r}{2c}&-\frac{\mathcal{ E}+\Delta-V(r)}{v_{F}}\\
\frac{\mathcal{ E}-\Delta-V(r)}{v_{F}}&-\frac{J}{r}-
\frac{e H r}{2c}\end{array}\right] \\
& = \frac{\hbar}{l}\left[\begin{array}{cc}\frac{j}{\rho}+\frac{\rho}{2}&\beta\rho-\epsilon-\delta\\
-(\beta\rho-\epsilon+\delta)& -\frac{j}{\rho}-\frac{\rho}{2} \end{array}\right].
\end{split}
\end{equation}
Here we denoted $J = \hbar j$ in the first line and introduced
the dimensionless variable $\rho = r/l$ with $l = \sqrt{\hbar c/(e H)}$
being the magnetic length in the second one.
We also introduced the dimensionless energy $\epsilon = l \mathcal{ E}/(\hbar v_F)$, and mass (gap) $\delta = l \Delta/(\hbar v_F)$.
The key dimensionless parameter, $\beta = c E/(v_F H)$, characterizes the relative strength of the electric field
compared to the magnetic field. In this paper, we limit our analysis to the case where $|\beta| \leq 1/2$
and exclude the pair creation regime from consideration.

\section{WKB method for Dirac fermions}
\label{sec:WKB}

Applications of the WKB method to the Dirac equation were often based on reducing it to a
Schr\"{o}dinger-like equation for one of components
with an effective potential (see, for example, Ref.~\cite{Zeldovich1972UFN}). The problem
with this approach is the appearance in an effective potential of additional (apparent) singularities.
Additionally, a Schr\"{o}dinger-like
equation for a single component may turn out  a differential equation of higher than second order, for which the WKB method
(particularly Bohr-Sommerfeld-like quantization) is not well developed.
Therefore, in this paper, we adopt an alternative approach by expanding the initial system
(\ref{WKB-Dirac}) in powers of $\hbar$.

As evident from the first line of Eq.~(\ref{matrix-D}), the system of equations (\ref{WKB-Dirac}) contains a small parameter $\hbar$,
allowing the use of the standard asymptotic scheme for solving systems of linear differential equations
\cite{Pauli1932HPA,Rubinow1963PRev,Popov1978SNP,Lazur2005TMP,Zhang2012PRB}, 
which forms the foundation of the WKB method. In the WKB expansion, energy and total angular momentum  $J=\hbar j$ are treated as independent of $\hbar$
since they are conserved physical quantities. Consequently, the matrix $M$ is independent of $\hbar$. Thus, one can seek
solution in the following form:
\begin{equation}
\label{WKB-series}
\Phi(r)=\exp \left[ \frac{i}{\hbar} S(r)\right] \Psi(r),\quad \Psi(r)=\sum\limits_{n=0}^\infty (-i\hbar)^n \Psi^{(n)}(r).
\end{equation}
Then the equation for the spinor $\Psi(r)$ acquires the form:
\begin{equation}
\label{eq:Psi}
(M_0+\hbar M_1)\Psi=0,
\end{equation}
where the matrices $M_0, M_1$ are given by
\begin{equation}
\begin{split}
& M_0=\left(\begin{array}{cc}a(r)-iS'(r)&-b(r)\\ b_1(r)&-a(r)-iS'(r) \end{array}\right),\\
&M_1=-\left(\begin{array}{cc}1&0\\0&1
\end{array}\right)\frac{d}{dr}
\end{split}
\end{equation}
with the functions
\begin{equation}
\label{a-b-b1}
\begin{split}
a(r)=\frac{J}{r}+\frac{e H r}{2c},\quad b(r) & =\frac{\mathcal{E}+\Delta-V(r)}{v_F},\\
b_1(r) & =\frac{\mathcal{E}-\Delta-V(r)}{v_F}.
\end{split}
\end{equation}
Substituting the asymptotic series for $\Psi(r)$  in powers of $\hbar$, as given by Eq.~(\ref{WKB-series}),
into Eq.~(\ref{eq:Psi}) and equating the coefficients of equal powers of $\hbar$,
yields the first two equations for $\Psi^{(0)}$ and $\Psi^{(1)}$:
\begin{subequations}
\label{Psi}
\begin{align}
\label{Psi_zero}
& M_0\Psi^{(0)}=0,\qquad \mbox{or}\qquad M\Psi^{(0)}=iS'\Psi^{(0)}, \\
\label{Psi_1}
& M_1\Psi^{(0)}-iM_0\Psi^{(1)}=0.
\end{align}
\end{subequations}

A nontrivial solution for $\Psi^{(0)}$ exists if $\mbox{det} M_0=0$, which yields:
$S'_{\pm}(r)=\pm p(r)$, where
\begin{equation}
\label{p(r)}
\begin{split}
p(r)&= \sqrt{b(r)b_1(r)-a^2(r)}\\
& = \sqrt{\frac{(\mathcal{E}-V)^2-\Delta^2}{v_F^2}-
\left(\frac{J}{r}+\frac{e  H r}{2c}\right)^2}.
\end{split}
\end{equation}
is the classical local momentum for the radial motion of a particle.
The right and left null vectors of the matrix $M_0$, corresponding to eigenvalues $i S'(r)=\pm i p(r)$
of the matrix $M$ are defined as follows, respectively:
\begin{subequations}
\label{right-left-def}
\begin{align}
\label{right-def}
M_0 \Psi_{\pm}^{(0)}(r) =0, \\
\label{left-def}
\tilde{\Psi}_{\pm}^{(0)}(r)M_0=0.
\end{align}
\end{subequations}
Accordingly, we find that the right null vectors can be written in the
two forms
\begin{equation}
\label{right-spinors}
\begin{split}
\Psi_{\pm}^{(0)}(r)& =C^{\pm}_R(r)\left(\begin{array}{c}b(r)\\a(r)\mp ip(r)\end{array}\right) \\
& =D^{\pm}_R(r)\left(\begin{array}{c}
a(r)\pm i p(r)\\b_1(r)\end{array}\right),
\end{split}
\end{equation}
and left null vectors are
\begin{equation}
\begin{split}
\tilde{\Psi}_{\pm}^{(0)}(r) & =C^{\pm}_L(r)\left(-b_1(r), a(r)\mp ip(r)\right)\\
& = D^{\pm}_L(r)\left( a(r)\pm ip(r),-b(r)\right).
\end{split}
\end{equation}
Here the normalization functions $C^{\pm}_{L,R}(r)$, $D^{\pm}_{L,R}(r)$ will be determined in what follows.
Note that since the matrix $M_0$ is not symmetric its left eigenvectors do not coincide with its right eigenvectors:
$\tilde{\Psi}_{\pm}^{(0)}(r)\neq {\Psi_{\pm}^{(0)}}^T(r)$. It is easy to check that left and right eigenvectors are orthogonal for different eigenvalues,
$\tilde{\Psi}_{\pm}^{(0)}(r)\Psi_{\mp}^{(0)}(r)=0$.
Undetermined functions $C^{\pm}_R(r)$ and $D^{\pm}_{R}(r)$ are found from Eq.~(\ref{Psi_1}) if we set $\Psi^{(0)}=\Psi^{(0)}_{\pm}$
and multiply from the left by the row vector $\tilde{\Psi}_{\pm}^{(0)}(r)$. Then, in view of Eq.~(\ref{left-def}), the second term of this
equation has to be zero, and we obtain an equation for $C^{\pm}_R(r)$ in the following form
\begin{equation}
\label{Psi-0-eq}
\tilde{\Psi}_{\pm}^{(0)}(r)M_1\Psi^{(0)}_{\pm}=0.
\end{equation}
The equations for the functions $C^{\pm}_{L}(r)$ and $D^{\pm}_{L}(r)$ can be found in the same fashion.

The explicit form of the function $C^{\pm}_{R}(r)$ is given by Eq.~(\ref{C_R}) from Appendix~\ref{sec:Appendix-WKB-wave}
and the corresponding spinor wave function $\Psi^{(0)}_{\pm}(r)$ is represented by
Eq.~(\ref{gapped_graphene-WKB_solutions}) [see also Eq.~(\ref{gapped_graphene-WKB_solutions-2})].
It is convenient to represent the solution of Eq.~(\ref{gapped_graphene-WKB_solutions}) with the phase explicitly extracted.
To achieve this, we write:
\begin{equation}
\begin{split}
& p(r)\pm ia(r)=\sqrt{p^2(r)+a^2(r)}e^{\pm 2i\varphi(r)},\\
& \varphi(r)=\frac{1}{2}\arctan\frac{a(r)}{p(r)}
\end{split}
\end{equation}
and
\begin{equation}
p^2(r)+a^2(r)=b(r)b_1(r)=\frac{(\mathcal{E}-V(r))^2-\Delta^2}{v_F^2}.
\end{equation}
Thus, we obtain the following solution in the classically allowed region
\begin{equation}
\label{Psi-WKB-nonzeroDelta}
\begin{split}
 \Phi^{(0)}_{\pm}(r)  =  &\frac{\tilde{C}^{\pm}}{[p(r)|b(r)|]^{1/2}}
\exp\left(\pm \frac{i}{\hbar}\int^rp(r)dr \pm i\gamma(r)\right) \\
& \times \left(\begin{array}{c}b(r)e^{\mp i\varphi(r)}\\ \mp i\sqrt{b(r)b_1(r)}\, e^{\pm i\varphi(r)}\end{array}\right).
\end{split}
\end{equation}
Here $\gamma(r)$ is the geometrical phase defined by Eq.~(\ref{gamma-r-1})
which can be rewritten as follows
\begin{equation}
\label{gamma-r}
\gamma(r)= \frac{\Delta}{2}\int^r\frac{dr V'(r)a(r)}{p(r)[(E-V(r))^2-\Delta^2]}.
\end{equation}

\section{Generalized Bohr-Sommerfeld quantization condition}
\label{sec:Bohr}

As mentioned in Introduction, we  derive the BS-like quantization condition
using the method proposed by Wentzel \cite{Wentzel1926ZP} and further developed in \cite{Dunham1932PhysRev}
(see also \cite{Froman1977JMP,Parisi1979book}).
Wentzel's original consideration is based on the assumption that the wave functions in the classically
allowed region are known to have $n_{\mathrm{BS}}$ nodes.

In the complex plane, the number of zeros $n$ of an analytic hence single valued function $\psi(z)$
inside a contour $C$ traversed in an anticlockwise direction is determined by the argument principle:
\begin{equation}
\label{argument_principle}
\oint_C\frac{\psi'(z)}{\psi(z)}dz=  \oint_C[\ln \psi (z) ]'dr = 2\pi i n.
\end{equation}

Let us apply the argument principle to the wave function $\Phi^{(0)}_{+}(r)$.
For the positive energy, the function $b(r)$ defined by Eq.~(\ref{a-b-b1}) has no zeros.
Therefore, for the upper spinor component $\Phi^{(0)}_{+,1}(r)$, we can use the solution given by Eq.~(\ref{Psi-WKB-nonzeroDelta}).
For the negative energy, the function $b_1(r)$ (see Eq.~(\ref{a-b-b1}) has no zeros. Therefore,
for the lower spinor component, it is convenient to use
the solution in the form presented by Eq.~(\ref{gapped_graphene-WKB_solutions-2}).
Thus, we arrive at the following quantization rule:
\begin{equation}
\label{BS1}
\begin{split}
  i & \oint_C dr \left[  \frac{p(r)}{\hbar} + \gamma^\prime(r) -\frac{1}{2}\left(\arctan\frac{a(r)}{p(r)}\right)' \right]  \\
 - \frac{1}{2} & \oint_C dr  \frac{(p^2(r))'}{2p^2(r)}  =2 \pi i n_{\mathrm{BS}},
\end{split}
\end{equation}
where $n_{\mathrm{BS}}=0,1,2,\ldots$ is the BS quantum number and
the term  $\gamma^\prime(r)$ is the derivative of the geometrical phase (\ref{gamma-r}), which accounts for the finiteness of $\Delta$.
Note that for the lower spinor component $\Phi^{(0)}_{+,2}(r)$, the quantization rule is identical, except for the opposite sign
in front of the term involving $\arctan$. Eq.~(\ref{BS1}) can be regarded as the condition
of single valuedness of the WKB wave function \cite{Wentzel1926ZP,Dunham1932PhysRev,Zhang2012PRB}.

\begin{figure}[!h]
\includegraphics[width=.45\textwidth]{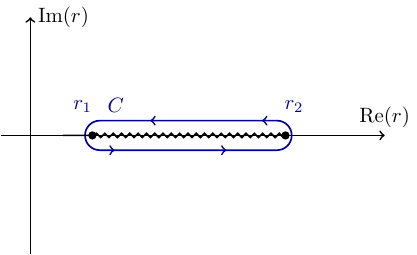}
\caption{The contour of integration in the $r$-plane.
}
\label{fig-contour}
\end{figure}

In the complex $r$-plane, we assume the presence of a branch cut between the two classical turning points
$r_1$ and $r_2$.  We choose $p(r)$  to be positive on the upper side of the branch cut (see Fig.~\ref{fig-contour}).
In turn, the function $p^2(r)$ has two simple roots, leading to the following:
\begin{equation}
\frac{1}{2}\oint_C\frac{(p^2(r))'}{2p^2(r)}dr=\pi i.
\end{equation}
Thus we obtain from Eq.~(\ref{BS1}) the following form of the quantization condition
\begin{equation}
\label{BS2}
\frac{1}{\hbar}\int\limits_{r_1}^{r_2}dr p(r) + \alpha
=\pi\left (n_{\mathrm{BS}}+\frac{1}{2} \right),
\end{equation}
where the phase
\begin{equation}
\label{alpha}
\begin{split}
\alpha &=\int\limits_{r_1}^{r_2}dr \left[\frac{\Delta}{2}\frac{ V'(r)a(r)}{p(r)[(E-V(r))^2-\Delta^2]}\right.\\
&\left.\mp \frac{1}{2}\left(\arctan\frac{a(r)}{p(r)}\right)' \right]dr
\end{split}
\end{equation}
describes the contribution of the spinor part of the wave function (signs $\mp$ correspond to the signs of phases in upper and lower
components of $\Phi^{(0)}_{+}(r)$).

Finally, one should analyze the integral with $\arctan$ in Eq.~(\ref{alpha}).
For positive $j$, the function $a(r)$ defined by Eq.~(\ref{a-b-b1})
is always positive and the integral with $\arctan $ in Eq.~(\ref{alpha})
equals to zero. For negative $j$, $a(r)$ is monotonically increasing function, which changes sign at the point
$r_0=l\sqrt{2|j|}$ in the classically allowed region.  One can justify that
for the considered model  with $p(r)$ given by Eq.~(\ref{p(r)}) as well as for the model
with the Coulomb form of $V(r)$ (see Ref.~\cite{Zhang2012PRB}) the point $r_0$ is inside the interval: $r_1 < r_0 < r_2$.
Hence we obtain
\begin{align}
\int\limits_{r_1}^{r_2}dr\left(\arctan\frac{a(r)}{p(r)}\right)' =  \pi \theta (-j) .
\end{align}
Thus, we arrive at the final form of the generalized BS quantization condition:
\begin{align}
\label{BS-final}
&\int\limits_{r_1}^{r_2}dr \left[ p(r) + \frac{\hbar\Delta}{2}\frac{ V'(r)a(r)}{p(r)[(E-V(r))^2-\Delta^2]} \right]\nonumber\\
&= \pi \hbar \left (n_{\mathrm{BS}}+\frac{1 \pm \theta(-j)}{2} \right)
\end{align}
with $n_{\mathrm{BS}}=0,1,2,\ldots$. The form of Eq.~(\ref{BS-final}) with the minus sign  before theta-function is more general, because its
right hand side starts from the zero value. Thus, in what follows we will use the form with $1 - \theta(-j)$.
We emphasize that since $\gamma(r)$ in Eq.~(\ref{gamma-r}) depends on $\Delta$ rather than $\Delta^2$,
the inclusion of this term in Eq.~(\ref{BS-final}) renders the spectra valley-dependent. This is because, as noted below Eq.~(\ref{Dirac-eq-2*2}),
the results for $\mathbf{K}_-$ valley are obtained by reversing sign of $\Delta$.
When $\Delta = 0$ and/or $V^\prime (r) =0$, hence, $\gamma (r) =0$, Eq.~(\ref{BS-final}) reduces to the quantization condition
considered in Refs.~\cite{Zhang2012PRB,Nimyi2022PRB}.

\subsection{WKB approximation in the absence of electric field}
\label{sec:E=0}

To illustrate how one should deal with the quantum numbers in the symmetric gauge, we
briefly consider the problem in the absence of an electric field, $E =0$, when it is
exactly solvable (see Appendix~D of
Ref.~\cite{Gusynin2006PRB}, where the final result is written in
the form identical to the spectrum obtained
in the Landau gauge).

In this case, the integral on the
left hand side of the BS-like quantization condition Eq.~(\ref{BS-final})
can be calculated:
\begin{equation}
\label{int-E=0}
\int\limits_{r_1}^{r_2}dr p(r) =  \frac{\hbar \pi}{2}
(\epsilon^2 - \delta^2 - |j| - j),
\end{equation}
where the dimensionless energy $\epsilon$ and gap $\delta$ were defined below Eq.~(\ref{matrix-D}).
Solving Eq.~(\ref{BS-final}) for the energy $\epsilon$, one recovers the
well-known Landau-level spectrum,
\begin{equation}
\label{LLspectrum-magfield}
\mathcal{ E}=\pm \sqrt{\Delta^2 + (2n_{\mathrm{BS}} + j+|j|+1-\theta(-j)) e H \hbar v_F^2/c}
\end{equation}
with $n_{\mathrm{BS}}=0,1,2,\ldots$.
Except for the asymmetric lowest Landau level,
this result  agrees perfectly with the exact solution
of the Dirac equation in Ref.~\cite{Gusynin2006PRB}.

For $j<0$, the spectrum Eq.~(\ref{LLspectrum-magfield})
is consistent with  the result in the Landau gauge if one identifies
the quantum number $n_{\mathrm{BS}}$ and  the Landau level index $n = n_{\mathrm{BS}}$.
To reproduce the spectrum for the $j>0$ case, one should relabel
the quantum numbers
$2n =2 n_{\mathrm{BS}} + 2 j +1$, so  $n$ corresponds to
the Landau-level index
and $\mathcal{ E}=\pm \sqrt{\Delta^2 + 2n  e H \hbar v_F^2/c}$
with $n=1,2, \ldots $ and $1/2 \leq j\leq n -1/2$.
Thus, one observes that in the absence of an electric field,
the WKB method reproduces well-known exact results (see, for example, Ref.~\cite{Zhang2012PRB}).

\section{Results}
\label{sec:results}

It is convenient to rewrite Eq.~(\ref{BS-final}) using the dimensionless variable
$\rho = r/l$. Denoting $\rho_{1,2} = r_{1,2}/l$ and
introducing the dimensionless momentum $p(\rho)$ by
\begin{equation}
p(r)=\frac{\hbar}{l} p (\rho), \qquad p (\rho) =
\sqrt{(\epsilon-\beta\rho)^2-\delta^2-\left(\frac{j}{\rho}+\frac{\rho}{2}\right)^2},
\end{equation}
we write the quantization condition Eq.~(\ref{BS-final}) as follows
\begin{equation}
\label{BS-final-dimensionless}
\begin{split}
& \int\limits_{\rho_1}^{\rho_1}d\rho p(\rho )+\frac{\beta\delta}{2}
\int\limits_{\rho_1}^{\rho_2}\frac{d\rho(j/\rho + \rho/2)}{p(\rho)
[(\epsilon-\beta\rho)^2-\delta^2]} \\
& =\pi\left(n_{\mathrm{BS}}+\frac{1-\theta(-j)}{2}\right).
\end{split}
\end{equation}

\begin{figure*}[!htb]
\includegraphics[width=.95\textwidth]{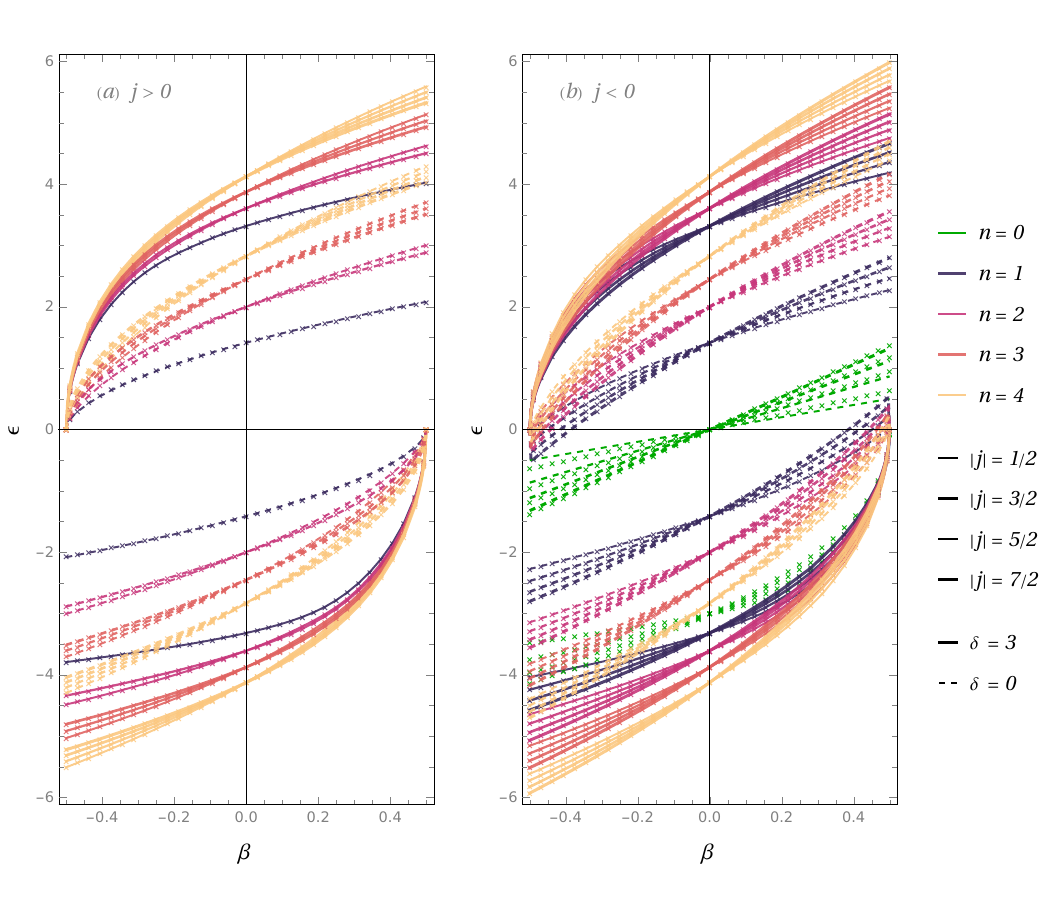}
\caption{WKB  and exact diagonalization (ED) spectra in units
$\epsilon_0 = \hbar v_F/l$ versus electric field in terms of $\beta = c E/(v_F H)$
for the finite gap,
$\delta =  l \Delta/(\hbar v_F) $, case.
The solid lines are the solutions for
$\delta =3$ and  the dashed lines
are for comparison $\delta =0$.
(a) Positive $j$.
(b) Negative $j$.
In both panels, solid and dashed lines show the results of the WKB approximation and crosses
show the results calculated using ED.
The Landau levels with $n=0, 1,2,3,4$ are shown by the green,
dark purple, purple lines, red and orange lines, respectively.
The levels with $|j| =1/2, 3/2,5/2,7/2$ are marked by the increasing
thickness of the lines.
}
\label{fig-spectra-gapped}
\end{figure*}

The BS quantization condition Eq.~(\ref{BS-final}) [see also Eq.~(\ref{BS-final-dimensionless})] represents the
transcendental  WKB equation for the energy spectrum.
For $\Delta =0$, the integral on the left-hand side of the equation (\ref{BS-final-dimensionless})
can be expressed in terms of complete Legendre elliptic integrals
\cite{Nimyi2022PRB}. However, in the general case of a finite $\Delta$, it has to be evaluated numerically.
The numerical solution, which describes the dependence of the energy $\mathcal{E} $
(in units of  $\epsilon_0 = \hbar v_F/l$) on the electric field in terms of the dimensionless parameter $\beta$,
is shown in Fig.~\ref{fig-spectra-gapped}.
For clarity, the cases
$j>0$ and $j <0$  are presented separately in the left (a) and right (b) panels, respectively.
For $j >0$, the Landau levels have been relabeled as described below Eq.~(\ref{LLspectrum-magfield}),
so the notation $n=1,2,3, 4$ corresponds to the Landau level index. The lowest level, $n=0$, exists only for $j<0$.
First, we observe that the presence of a finite electric field lifts the degeneracy of the Landau levels with different
total angular momenta $|j| =1/2,3/2,5/2, 7/2$ , which are indicated by progressively increasing line thickness.

To verify the accuracy of the WKB approximation, we employ a numerical method involving discretization
followed by exact diagonalization of the the Hamiltonian derived from Eq.~(\ref{matrix-D}).
To implement the discretization method, the {\large K}WANT \cite{kwant} Python package was used. With this package,
we build the Hamiltonian of a 2000-site-long chain and then numerically diagonalize it.
The details are discussed in Ref.~\cite{Nimyi2022PRB}, and it is important to note that this method reveals
a fermionic doubling effect
\cite{Nielsen1981NP} for a non-zero gap, resulting in spectrum doubling.
In Ref.~\cite{Nimyi2022PRB}, we investigated this point using the shooting method,
which enabled us to numerically solve the corresponding equations for each $\mathbf{K}_{\pm}$ separately.
This approach confirmed that the diagonalization method combines the results for both valleys.
Since the exact diagonalization method is more efficient in terms of speed and precision compared to the shooting method,
we relied on the former approach in this work.

The results of the fully numerical method
for the dependence of the energy $\mathcal{E} $
(in units $\epsilon_0 = \hbar v_F/l$) versus electric field in terms of
the dimensionless parameter $\beta$ in the case of a finite gap $\delta$
are shown in Fig.~\ref{fig-spectra-gapped} by crosses.

The comparison with WKB (solid lines in Fig.~\ref{fig-spectra-gapped}) shows that the WKB approximation is in a good
quantitative agreement with the numerical calculations for the whole range of the values of $\beta$, $j$, and $n \neq 0$.
The WKB approximation breaks for the lowest, $n = 0$, Landau level with $\delta \neq 0$, so in
Fig.~\ref{fig-spectra-gapped}~(b) we show  only the exact diagonalization results.
While the direct substitution of the numerical diagonalization results
in Eq.~(\ref{BS-final-dimensionless}) for the $n \neq 0$
levels shows that it is satisfied up to $10^{-6}$, this is not the case of the $n=0$ level.

In  Ref.~\cite{Nimyi2022PRB} we have considered in detail the behavior of the energy levels in the
vicinity of the points $|\beta| = 1/2$.
For $\delta = 0$, the behavior of levels with $j < 0$ is drastically different from that
in the $j > 0$ case, where the Landau level collapse was found.
The gapped case for $j <0$ turned out to be even more
complicated and depending on the relationship between $|j|$ and $\delta^2$.
In Fig.~\ref{fig-spectra-gapped}~(b) the shown levels do collapse for $\delta =3$
and do not collapse for $\delta=0$.
One can find a complete discussion of the Landau level collapse in the gapped case in  Ref.~\cite{Nimyi2022PRB}.

\section{Conclusion}
\label{sec:conclusion}

The appearance of a geometric phase in WKB solutions of multicomponent wave fields is a general fact in many cases and leads
to a generalization of the Bohr-Sommerfeld quantization condition for scalar fields
\cite{Littlejohn1991PRA}. In present paper, we applied the WKB method to the system of radial equations in gapped graphene
under magnetic and radial in-plane electric fields.
Using the WKB approximation, we
demonstrated that for gapped graphene the generalized quantization condition (\ref{BS-final})
for eigenenergies includes an additional
valley-dependent geometrical phase $\gamma(r)$ given by Eq.~(\ref{gamma-r}).
When the gap
$\Delta = 0$ and/or the potential does not depend on the radial coordinate, $V^\prime (r) =0$,
this phase is absent, so Eq.~(\ref{BS-final}) reduces to the quantization condition
considered in Refs.~\cite{Zhang2012PRB,Nimyi2022PRB}.

However, this term turns out to be crucial for reaching an agrement with
the exact diagonalization numerical results. It would also be interesting to consider
a possibility of an experimental observation of the effects related to
the nonzero geometrical phase. As mentioned in the Introduction, the collapse of Landau levels
has already been observed experimentally \cite{Singh2009PRB,Gu2011PRL}.
Although these observations were conducted in a rectangular geometry, there are no obstacles to replicating
them in the circular Corbino geometry explored in this study.
We also note that a global $A/B$ sublattice asymmetry gap of $2 \Delta \sim \SI{350}{K}$ can be induced in graphene
(see e.g. \cite{Gorbachev2014Science})
when it is placed on hexagonal boron nitride (G/hBN) with the crystallographic axes of graphene and hBN aligned.
Another unexplored option is to investigate the connection between the phenomenon of collapsing Landau levels and
studies involving tilted Dirac cones in a strong magnetic field \cite{Goerbig2009EPL}. It is possible that systems
with tilted Dirac cones are easier to realize experimentally compared to creating crossed field configurations.

\begin{acknowledgments}

We are grateful to the Armed Forces of Ukraine for providing security to perform this work.
One of us (S.~G.~Sh.) regards academician Vadym Loktev as a mentor who also played a pivotal role in his early career,
particularly during  complicated times.
I.O.N., S.G.Sh and V.P.G. acknowledge the support from the
National Academy of Sciences of Ukraine Projects No.~0122U000887
and No.~0121U109612, respectively.
The work was also supported by the Simons Foundation (U.S.).


\end{acknowledgments}	

\appendix

\section{WKB wave functions}
\label{sec:Appendix-WKB-wave}

\begin{widetext}

Starting from Eq.~(\ref{Psi-0-eq}),
we derive, for example, the following equation for the functions $C^{\pm}_R(r)$:
\begin{equation}
\frac{\partial_r C^{\pm}_R}{C^{\pm}_R}=-\frac{1}{2}\frac{d}{dr}\ln[-b b_1+(a\mp ip)^2]-\frac{1}{2}\frac{b b'_1-b'b_1}{-b b_1+(a\mp ip)^2},
\end{equation}
where, for brevity, the arguments of the functions $a(r)$, $b(r)$, $b_1(r)$, and $p(r)$ were omitted.
The solution to this equation is expressed as:
\begin{equation}
\label{C_R}
C^{\pm}_R(r)= \frac{\tilde{C}^{\pm}}{[p(r)(p(r)\pm ia(r))]^{1/2}}\exp\left(-\int^rdr\frac{\Delta V'(r)}{2v_F^2p(r)(p(r)\pm ia(r))}\right),
\end{equation}
where $\tilde{C}^{\pm}$ are constants.
Finally, we obtain the following WKB solutions
\begin{equation}
\label{gapped_graphene-WKB_solutions}
\Phi^{(0)}_{\pm}(r)=\frac{\tilde{C}^{\pm}}{[p(r)(p(r)\pm ia(r))]^{1/2}}\exp\left(\pm \frac{i}{\hbar}\int^rp(r)dr+\Gamma(r)\right)
\left(\begin{array}{c}b(r)\\a(r)\mp ip(r)\end{array}\right),
\end{equation}
where the function
\begin{equation}
\label{Gamma}
\Gamma(r)=-\int^r\frac{\Delta V'(r)dr}{2v_F^2p(r)(p(r)\pm ia(r))}.
\end{equation}
The last expression can be written in the following form
\begin{equation}
\Gamma(r)=-\frac{1}{4}\ln \frac{E-V(r)+\Delta}{E-V(r)-\Delta}\pm i\gamma(r)=\frac{1}{4}\ln \frac{b(r)}{b_1(r)}\pm i\gamma(r),
\end{equation}
where we introduce the geometrical phase
\begin{equation}
\label{gamma-r-1}
\gamma(r)=\frac{1}{4}\int^r\frac{dr V'(r)a(r)}{p(r)}\left[\frac{1}{E-V(r)-\Delta}- \frac{1}{E-V(r)+\Delta}\right].
\end{equation}

We can obtain slightly different form for the spinors $\Phi^{(0)}_{\pm}(r)$ using the second line in
Eq.~(\ref{right-spinors}) for the right spinor:
\begin{equation}
\label{gapped_graphene-WKB_solutions-2}
\begin{split}
 \Phi^{(0)}_{\pm}(r)  =  &\frac{\tilde{C}^{\pm}}{[p(r)|b_1(r)|]^{1/2}}
\exp\left(\pm \frac{i}{\hbar}\int^rp(r)dr \pm i\gamma(r)\right) \\
& \times \left(\begin{array}{c} \pm i\sqrt{b(r)b_1(r)}\, e^{\mp i\varphi(r)}\\ b_1(r)e^{\pm i\varphi(r)}\\\end{array}\right).
\end{split}
\end{equation}
Different forms are used in order to avoid a singularity of the wave function at the point where $b(r)=0$ or $b_1(r)=0$ within
classically allowed region.  Moreover, in classically forbidden regions, where after analytic continuation $p(r)\rightarrow \pm iq(r)$,
the choice of forms depends on the sign of $a(r)$ to avoid singularity.
\end{widetext}

\bibliography{WKB-collapse.bib}

\begin{thebibliography}{34}%
\makeatletter
\providecommand \@ifxundefined [1]{%
 \@ifx{#1\undefined}
}%
\providecommand \@ifnum [1]{%
 \ifnum #1\expandafter \@firstoftwo
 \else \expandafter \@secondoftwo
 \fi
}%
\providecommand \@ifx [1]{%
 \ifx #1\expandafter \@firstoftwo
 \else \expandafter \@secondoftwo
 \fi
}%
\providecommand \natexlab [1]{#1}%
\providecommand \enquote  [1]{``#1''}%
\providecommand \bibnamefont  [1]{#1}%
\providecommand \bibfnamefont [1]{#1}%
\providecommand \citenamefont [1]{#1}%
\providecommand \href@noop [0]{\@secondoftwo}%
\providecommand \href [0]{\begingroup \@sanitize@url \@href}%
\providecommand \@href[1]{\@@startlink{#1}\@@href}%
\providecommand \@@href[1]{\endgroup#1\@@endlink}%
\providecommand \@sanitize@url [0]{\catcode `\\12\catcode `\$12\catcode
  `\&12\catcode `\#12\catcode `\^12\catcode `\_12\catcode `\%12\relax}%
\providecommand \@@startlink[1]{}%
\providecommand \@@endlink[0]{}%
\providecommand \url  [0]{\begingroup\@sanitize@url \@url }%
\providecommand \@url [1]{\endgroup\@href {#1}{\urlprefix }}%
\providecommand \urlprefix  [0]{URL }%
\providecommand \Eprint [0]{\href }%
\providecommand \doibase [0]{https://doi.org/}%
\providecommand \selectlanguage [0]{\@gobble}%
\providecommand \bibinfo  [0]{\@secondoftwo}%
\providecommand \bibfield  [0]{\@secondoftwo}%
\providecommand \translation [1]{[#1]}%
\providecommand \BibitemOpen [0]{}%
\providecommand \bibitemStop [0]{}%
\providecommand \bibitemNoStop [0]{.\EOS\space}%
\providecommand \EOS [0]{\spacefactor3000\relax}%
\providecommand \BibitemShut  [1]{\csname bibitem#1\endcsname}%
\let\auto@bib@innerbib\@empty
\bibitem [{\citenamefont {Nimyi}\ \emph {et~al.}(2022)\citenamefont {Nimyi},
  \citenamefont {K\"onye}, \citenamefont {Sharapov},\ and\ \citenamefont
  {Gusynin}}]{Nimyi2022PRB}%
  \BibitemOpen
  \bibfield  {author} {\bibinfo {author} {\bibfnamefont {I.~O.}\ \bibnamefont
  {Nimyi}}, \bibinfo {author} {\bibfnamefont {V.}~\bibnamefont {K\"onye}},
  \bibinfo {author} {\bibfnamefont {S.~G.}\ \bibnamefont {Sharapov}},\ and\
  \bibinfo {author} {\bibfnamefont {V.~P.}\ \bibnamefont {Gusynin}},\
  }\bibfield  {title} {\bibinfo {title} {Landau level collapse in graphene in
  the presence of in-plane radial electric and perpendicular magnetic fields},\
  }\href {https://doi.org/10.1103/PhysRevB.106.085401} {\bibfield  {journal}
  {\bibinfo  {journal} {Phys. Rev. B}\ }\textbf {\bibinfo {volume} {106}},\
  \bibinfo {pages} {085401} (\bibinfo {year} {2022})}\BibitemShut {NoStop}%
\bibitem [{\citenamefont {Singh}\ and\ \citenamefont
  {Deshmukh}(2009)}]{Singh2009PRB}%
  \BibitemOpen
  \bibfield  {author} {\bibinfo {author} {\bibfnamefont {V.}~\bibnamefont
  {Singh}}\ and\ \bibinfo {author} {\bibfnamefont {M.~M.}\ \bibnamefont
  {Deshmukh}},\ }\bibfield  {title} {\bibinfo {title} {Nonequilibrium breakdown
  of quantum {Hall} state in graphene},\ }\href
  {https://doi.org/10.1103/PhysRevB.80.081404} {\bibfield  {journal} {\bibinfo
  {journal} {Phys. Rev. B}\ }\textbf {\bibinfo {volume} {80}},\ \bibinfo
  {pages} {081404} (\bibinfo {year} {2009})}\BibitemShut {NoStop}%
\bibitem [{\citenamefont {Gu}\ \emph {et~al.}(2011)\citenamefont {Gu},
  \citenamefont {Rudner}, \citenamefont {Young}, \citenamefont {Kim},\ and\
  \citenamefont {Levitov}}]{Gu2011PRL}%
  \BibitemOpen
  \bibfield  {author} {\bibinfo {author} {\bibfnamefont {N.}~\bibnamefont
  {Gu}}, \bibinfo {author} {\bibfnamefont {M.}~\bibnamefont {Rudner}}, \bibinfo
  {author} {\bibfnamefont {A.}~\bibnamefont {Young}}, \bibinfo {author}
  {\bibfnamefont {P.}~\bibnamefont {Kim}},\ and\ \bibinfo {author}
  {\bibfnamefont {L.}~\bibnamefont {Levitov}},\ }\bibfield  {title} {\bibinfo
  {title} {Collapse of {Landau} levels in gated graphene structures},\ }\href
  {https://doi.org/10.1103/PhysRevLett.106.066601} {\bibfield  {journal}
  {\bibinfo  {journal} {Phys. Rev. Lett.}\ }\textbf {\bibinfo {volume} {106}},\
  \bibinfo {pages} {066601} (\bibinfo {year} {2011})}\BibitemShut {NoStop}%
\bibitem [{\citenamefont {Lukose}\ \emph {et~al.}(2007)\citenamefont {Lukose},
  \citenamefont {Shankar},\ and\ \citenamefont {Baskaran}}]{Lukose2007PRL}%
  \BibitemOpen
  \bibfield  {author} {\bibinfo {author} {\bibfnamefont {V.}~\bibnamefont
  {Lukose}}, \bibinfo {author} {\bibfnamefont {R.}~\bibnamefont {Shankar}},\
  and\ \bibinfo {author} {\bibfnamefont {G.}~\bibnamefont {Baskaran}},\
  }\bibfield  {title} {\bibinfo {title} {Novel electric field effects on
  {Landau} levels in graphene},\ }\href
  {https://doi.org/10.1103/PhysRevLett.98.116802} {\bibfield  {journal}
  {\bibinfo  {journal} {Phys. Rev. Lett.}\ }\textbf {\bibinfo {volume} {98}},\
  \bibinfo {pages} {116802} (\bibinfo {year} {2007})}\BibitemShut {NoStop}%
\bibitem [{\citenamefont {Peres}\ and\ \citenamefont
  {Castro}(2007)}]{Peres2007JPCM}%
  \BibitemOpen
  \bibfield  {author} {\bibinfo {author} {\bibfnamefont {N.~M.~R.}\
  \bibnamefont {Peres}}\ and\ \bibinfo {author} {\bibfnamefont {E.~V.}\
  \bibnamefont {Castro}},\ }\bibfield  {title} {\bibinfo {title} {Algebraic
  solution of a graphene layer in transverse electric and perpendicular
  magnetic fields},\ }\href@noop {} {\bibfield  {journal} {\bibinfo  {journal}
  {J. Phys. Condens. Matter}\ }\textbf {\bibinfo {volume} {19}},\ \bibinfo
  {pages} {406231} (\bibinfo {year} {2007})}\BibitemShut {NoStop}%
\bibitem [{\citenamefont {Ho}\ and\ \citenamefont
  {Khalilov}(2000)}]{Khalilov2000PRA}%
  \BibitemOpen
  \bibfield  {author} {\bibinfo {author} {\bibfnamefont {C.-L.}\ \bibnamefont
  {Ho}}\ and\ \bibinfo {author} {\bibfnamefont {V.~R.}\ \bibnamefont
  {Khalilov}},\ }\bibfield  {title} {\bibinfo {title} {Planar {Dirac} electron
  in {Coulomb} and magnetic fields},\ }\href
  {https://doi.org/10.1103/PhysRevA.61.032104} {\bibfield  {journal} {\bibinfo
  {journal} {Phys. Rev. A}\ }\textbf {\bibinfo {volume} {61}},\ \bibinfo
  {pages} {032104} (\bibinfo {year} {2000})}\BibitemShut {NoStop}%
\bibitem [{\citenamefont {Gamayun}\ \emph {et~al.}(2011)\citenamefont
  {Gamayun}, \citenamefont {Gorbar},\ and\ \citenamefont
  {Gusynin}}]{Gamayun2011PRB}%
  \BibitemOpen
  \bibfield  {author} {\bibinfo {author} {\bibfnamefont {O.~V.}\ \bibnamefont
  {Gamayun}}, \bibinfo {author} {\bibfnamefont {E.~V.}\ \bibnamefont
  {Gorbar}},\ and\ \bibinfo {author} {\bibfnamefont {V.~P.}\ \bibnamefont
  {Gusynin}},\ }\bibfield  {title} {\bibinfo {title} {Magnetic field driven
  instability of a charged center in graphene},\ }\href
  {https://doi.org/10.1103/PhysRevB.83.235104} {\bibfield  {journal} {\bibinfo
  {journal} {Phys. Rev. B}\ }\textbf {\bibinfo {volume} {83}},\ \bibinfo
  {pages} {235104} (\bibinfo {year} {2011})}\BibitemShut {NoStop}%
\bibitem [{\citenamefont {Zhang}\ \emph {et~al.}(2012)\citenamefont {Zhang},
  \citenamefont {Barlas},\ and\ \citenamefont {Yang}}]{Zhang2012PRB}%
  \BibitemOpen
  \bibfield  {author} {\bibinfo {author} {\bibfnamefont {Y.}~\bibnamefont
  {Zhang}}, \bibinfo {author} {\bibfnamefont {Y.}~\bibnamefont {Barlas}},\ and\
  \bibinfo {author} {\bibfnamefont {K.}~\bibnamefont {Yang}},\ }\bibfield
  {title} {\bibinfo {title} {Coulomb impurity under magnetic field in graphene:
  A semiclassical approach},\ }\href
  {https://doi.org/10.1103/PhysRevB.85.165423} {\bibfield  {journal} {\bibinfo
  {journal} {Phys. Rev. B}\ }\textbf {\bibinfo {volume} {85}},\ \bibinfo
  {pages} {165423} (\bibinfo {year} {2012})}\BibitemShut {NoStop}%
\bibitem [{\citenamefont {Sun}\ and\ \citenamefont {Zhu}(2014)}]{Sun2014PRB}%
  \BibitemOpen
  \bibfield  {author} {\bibinfo {author} {\bibfnamefont {S.}~\bibnamefont
  {Sun}}\ and\ \bibinfo {author} {\bibfnamefont {J.-L.}\ \bibnamefont {Zhu}},\
  }\bibfield  {title} {\bibinfo {title} {Impurity spectra of graphene under
  electric and magnetic fields},\ }\href
  {https://doi.org/10.1103/PhysRevB.89.155403} {\bibfield  {journal} {\bibinfo
  {journal} {Phys. Rev. B}\ }\textbf {\bibinfo {volume} {89}},\ \bibinfo
  {pages} {155403} (\bibinfo {year} {2014})}\BibitemShut {NoStop}%
\bibitem [{\citenamefont {Moldovan}\ \emph {et~al.}(2017)\citenamefont
  {Moldovan}, \citenamefont {Masir},\ and\ \citenamefont
  {Peeters}}]{Moldovan2017-2D}%
  \BibitemOpen
  \bibfield  {author} {\bibinfo {author} {\bibfnamefont {D.}~\bibnamefont
  {Moldovan}}, \bibinfo {author} {\bibfnamefont {M.~R.}\ \bibnamefont
  {Masir}},\ and\ \bibinfo {author} {\bibfnamefont {F.~M.}\ \bibnamefont
  {Peeters}},\ }\bibfield  {title} {\bibinfo {title} {Magnetic field dependence
  of the atomic collapse state in graphene},\ }\href@noop {} {\bibfield
  {journal} {\bibinfo  {journal} {2d Mater.}\ }\textbf {\bibinfo {volume}
  {5}},\ \bibinfo {pages} {015017} (\bibinfo {year} {2017})}\BibitemShut
  {NoStop}%
\bibitem [{\citenamefont {Gorbar}\ \emph {et~al.}(2018)\citenamefont {Gorbar},
  \citenamefont {Gusynin},\ and\ \citenamefont {Sobol}}]{Gorbar2018FNT}%
  \BibitemOpen
  \bibfield  {author} {\bibinfo {author} {\bibfnamefont {E.~V.}\ \bibnamefont
  {Gorbar}}, \bibinfo {author} {\bibfnamefont {V.~P.}\ \bibnamefont
  {Gusynin}},\ and\ \bibinfo {author} {\bibfnamefont {O.~O.}\ \bibnamefont
  {Sobol}},\ }\bibfield  {title} {\bibinfo {title} {Electron states in the
  field of charged impurities in two-dimensional dirac systems (review
  article)},\ }\href@noop {} {\bibfield  {journal} {\bibinfo  {journal} {Low
  Temp. Phys.}\ }\textbf {\bibinfo {volume} {44}},\ \bibinfo {pages} {371}
  (\bibinfo {year} {2018})}\BibitemShut {NoStop}%
\bibitem [{\citenamefont {Rodriguez-Nieva}\ and\ \citenamefont
  {Levitov}(2016)}]{Rodriguez-Nieva2016PRB}%
  \BibitemOpen
  \bibfield  {author} {\bibinfo {author} {\bibfnamefont {J.~F.}\ \bibnamefont
  {Rodriguez-Nieva}}\ and\ \bibinfo {author} {\bibfnamefont {L.~S.}\
  \bibnamefont {Levitov}},\ }\bibfield  {title} {\bibinfo {title} {Berry phase
  jumps and giant nonreciprocity in {Dirac} quantum dots},\ }\href
  {https://doi.org/10.1103/PhysRevB.94.235406} {\bibfield  {journal} {\bibinfo
  {journal} {Phys. Rev. B}\ }\textbf {\bibinfo {volume} {94}},\ \bibinfo
  {pages} {235406} (\bibinfo {year} {2016})}\BibitemShut {NoStop}%
\bibitem [{\citenamefont {Berry}\ and\ \citenamefont
  {Mount}(1972)}]{Berry1972RPP}%
  \BibitemOpen
  \bibfield  {author} {\bibinfo {author} {\bibfnamefont {M.~V.}\ \bibnamefont
  {Berry}}\ and\ \bibinfo {author} {\bibfnamefont {K.~E.}\ \bibnamefont
  {Mount}},\ }\bibfield  {title} {\bibinfo {title} {Semiclassical
  approximations in wave mechanics},\ }\href@noop {} {\bibfield  {journal}
  {\bibinfo  {journal} {Rep. Prog. Phys.}\ }\textbf {\bibinfo {volume} {35}},\
  \bibinfo {pages} {315} (\bibinfo {year} {1972})}\BibitemShut {NoStop}%
\bibitem [{\citenamefont {Kemble}(2005)}]{Kemble2005book}%
  \BibitemOpen
  \bibfield  {author} {\bibinfo {author} {\bibfnamefont {E.~C.}\ \bibnamefont
  {Kemble}},\ }\href
  {https://www.amazon.com/Fundamental-Principles-Quantum-Mechanics-Applications/dp/0486441539}
  {\emph {\bibinfo {title} {The fundamental principles of quantum mechanics:
  With elementary applications}}}\ (\bibinfo  {publisher} {Dover Phoenix
  Editions, Reissue Edition},\ \bibinfo {year} {2005})\BibitemShut {NoStop}%
\bibitem [{\citenamefont {Landau}\ and\ \citenamefont
  {Lifshitz}(1981)}]{Landau_vol3book}%
  \BibitemOpen
  \bibfield  {author} {\bibinfo {author} {\bibfnamefont {L.~D.}\ \bibnamefont
  {Landau}}\ and\ \bibinfo {author} {\bibfnamefont {E.~M.}\ \bibnamefont
  {Lifshitz}},\ }\href@noop {} {\emph {\bibinfo {title} {Quantum mechanics:
  Non-relativistic theory}}}\ (\bibinfo  {publisher} {Elsevier},\ \bibinfo
  {year} {1981})\BibitemShut {NoStop}%
\bibitem [{\citenamefont {Davydov}(1976)}]{Davydov.book}%
  \BibitemOpen
  \bibfield  {author} {\bibinfo {author} {\bibfnamefont {A.~S.}\ \bibnamefont
  {Davydov}},\ }\href@noop {} {\emph {\bibinfo {title} {Quantum mechanics}}},\
  \bibinfo {edition} {2nd}\ ed.,\ Monographs in Natural Philosophy\ (\bibinfo
  {publisher} {Pergamon Press},\ \bibinfo {address} {London, England},\
  \bibinfo {year} {1976})\BibitemShut {NoStop}%
\bibitem [{\citenamefont {Wentzel}(1926)}]{Wentzel1926ZP}%
  \BibitemOpen
  \bibfield  {author} {\bibinfo {author} {\bibfnamefont {G.}~\bibnamefont
  {Wentzel}},\ }\bibfield  {title} {\bibinfo {title} {Eine verallgemeinerung
  der quantenbedingungen f\"{u}r die zwecke der wellenmechanik},\ }\href@noop
  {} {\bibfield  {journal} {\bibinfo  {journal} {Z. Physik}\ }\textbf {\bibinfo
  {volume} {38}},\ \bibinfo {pages} {518} (\bibinfo {year} {1926})}\BibitemShut
  {NoStop}%
\bibitem [{\citenamefont {Dunham}(1932)}]{Dunham1932PhysRev}%
  \BibitemOpen
  \bibfield  {author} {\bibinfo {author} {\bibfnamefont {J.~L.}\ \bibnamefont
  {Dunham}},\ }\bibfield  {title} {\bibinfo {title} {The
  {Wentzel-Brillouin-Kramers} method of solving the wave equation},\ }\href
  {https://doi.org/10.1103/PhysRev.41.713} {\bibfield  {journal} {\bibinfo
  {journal} {Phys. Rev.}\ }\textbf {\bibinfo {volume} {41}},\ \bibinfo {pages}
  {713} (\bibinfo {year} {1932})}\BibitemShut {NoStop}%
\bibitem [{\citenamefont {Fr{\"o}man}\ and\ \citenamefont
  {Fr{\"o}man}(1977)}]{Froman1977JMP}%
  \BibitemOpen
  \bibfield  {author} {\bibinfo {author} {\bibfnamefont {N.}~\bibnamefont
  {Fr{\"o}man}}\ and\ \bibinfo {author} {\bibfnamefont {P.~O.}\ \bibnamefont
  {Fr{\"o}man}},\ }\bibfield  {title} {\bibinfo {title} {On {Wentzel's} proof
  of the quantization condition for a single-well potential},\ }\href@noop {}
  {\bibfield  {journal} {\bibinfo  {journal} {J. Math. Phys.}\ }\textbf
  {\bibinfo {volume} {18}},\ \bibinfo {pages} {96} (\bibinfo {year}
  {1977})}\BibitemShut {NoStop}%
\bibitem [{\citenamefont {Balian}\ \emph {et~al.}(1979)\citenamefont {Balian},
  \citenamefont {Parisi},\ and\ \citenamefont {Voros}}]{Parisi1979book}%
  \BibitemOpen
  \bibfield  {author} {\bibinfo {author} {\bibfnamefont {R.}~\bibnamefont
  {Balian}}, \bibinfo {author} {\bibfnamefont {G.}~\bibnamefont {Parisi}},\
  and\ \bibinfo {author} {\bibfnamefont {A.}~\bibnamefont {Voros}},\ }\bibfield
   {title} {\bibinfo {title} {Quartic oscillator},\ }in\ \href@noop {} {\emph
  {\bibinfo {booktitle} {Feynman Path Integrals}}},\ \bibinfo {editor} {edited
  by\ \bibinfo {editor} {\bibfnamefont {S.}~\bibnamefont {Albeverio}}, \bibinfo
  {editor} {\bibfnamefont {P.}~\bibnamefont {Combe}}, \bibinfo {editor}
  {\bibfnamefont {R.}~\bibnamefont {H{\o}egh-Krohn}}, \bibinfo {editor}
  {\bibfnamefont {G.}~\bibnamefont {Rideau}}, \bibinfo {editor} {\bibfnamefont
  {M.}~\bibnamefont {Sirugue-Collin}}, \bibinfo {editor} {\bibfnamefont
  {M.}~\bibnamefont {Sirugue}},\ and\ \bibinfo {editor} {\bibfnamefont
  {R.}~\bibnamefont {Stora}}}\ (\bibinfo  {publisher} {Springer Berlin
  Heidelberg},\ \bibinfo {address} {Berlin, Heidelberg},\ \bibinfo {year}
  {1979})\ pp.\ \bibinfo {pages} {337--360}\BibitemShut {NoStop}%
\bibitem [{\citenamefont {Korm\'anyos}\ \emph {et~al.}(2008)\citenamefont
  {Korm\'anyos}, \citenamefont {Rakyta}, \citenamefont {Oroszl\'any},\ and\
  \citenamefont {Cserti}}]{Kormanyos2008PRB}%
  \BibitemOpen
  \bibfield  {author} {\bibinfo {author} {\bibfnamefont {A.}~\bibnamefont
  {Korm\'anyos}}, \bibinfo {author} {\bibfnamefont {P.}~\bibnamefont {Rakyta}},
  \bibinfo {author} {\bibfnamefont {L.}~\bibnamefont {Oroszl\'any}},\ and\
  \bibinfo {author} {\bibfnamefont {J.}~\bibnamefont {Cserti}},\ }\bibfield
  {title} {\bibinfo {title} {Bound states in inhomogeneous magnetic field in
  graphene: Semiclassical approach},\ }\href
  {https://doi.org/10.1103/PhysRevB.78.045430} {\bibfield  {journal} {\bibinfo
  {journal} {Phys. Rev. B}\ }\textbf {\bibinfo {volume} {78}},\ \bibinfo
  {pages} {045430} (\bibinfo {year} {2008})}\BibitemShut {NoStop}%
\bibitem [{\citenamefont {Herasymchuk}\ \emph {et~al.}(2024)\citenamefont
  {Herasymchuk}, \citenamefont {Sharapov},\ and\ \citenamefont
  {Gusynin}}]{Herasymchuk2024PRB}%
  \BibitemOpen
  \bibfield  {author} {\bibinfo {author} {\bibfnamefont {A.~A.}\ \bibnamefont
  {Herasymchuk}}, \bibinfo {author} {\bibfnamefont {S.~G.}\ \bibnamefont
  {Sharapov}},\ and\ \bibinfo {author} {\bibfnamefont {V.~P.}\ \bibnamefont
  {Gusynin}},\ }\bibfield  {title} {\bibinfo {title} {Peculiarities of the
  {Landau} level collapse in graphene ribbons in crossed magnetic and in-plane
  electric fields},\ }\href {https://doi.org/10.1103/PhysRevB.110.125403}
  {\bibfield  {journal} {\bibinfo  {journal} {Phys. Rev. B}\ }\textbf {\bibinfo
  {volume} {110}},\ \bibinfo {pages} {125403} (\bibinfo {year}
  {2024})}\BibitemShut {NoStop}%
\bibitem [{\citenamefont {Gusynin}\ \emph {et~al.}(2007)\citenamefont
  {Gusynin}, \citenamefont {Sharapov},\ and\ \citenamefont
  {Carbotte}}]{Gusynin2007IJMPB}%
  \BibitemOpen
  \bibfield  {author} {\bibinfo {author} {\bibfnamefont {V.~P.}\ \bibnamefont
  {Gusynin}}, \bibinfo {author} {\bibfnamefont {S.~G.}\ \bibnamefont
  {Sharapov}},\ and\ \bibinfo {author} {\bibfnamefont {J.~P.}\ \bibnamefont
  {Carbotte}},\ }\bibfield  {title} {\bibinfo {title} {{AC conductivity of
  graphene: from tight-binding model to $2+1$-dimensional quantum
  electrodynamics}},\ }\href {https://doi.org/10.1142/S0217979207038022}
  {\bibfield  {journal} {\bibinfo  {journal} {Int. J. Mod. Phys. B}\ }\textbf
  {\bibinfo {volume} {21}},\ \bibinfo {pages} {4611 } (\bibinfo {year}
  {2007})}\BibitemShut {NoStop}%
\bibitem [{\citenamefont {Zeldovich}\ and\ \citenamefont
  {Popov}(1972)}]{Zeldovich1972UFN}%
  \BibitemOpen
  \bibfield  {author} {\bibinfo {author} {\bibfnamefont {Y.~B.}\ \bibnamefont
  {Zeldovich}}\ and\ \bibinfo {author} {\bibfnamefont {V.~S.}\ \bibnamefont
  {Popov}},\ }\bibfield  {title} {\bibinfo {title} {Electronic structure of
  superheavy atoms},\ }\href@noop {} {\bibfield  {journal} {\bibinfo  {journal}
  {Sov. Phys. Uspekhi}\ }\textbf {\bibinfo {volume} {14}},\ \bibinfo {pages}
  {673} (\bibinfo {year} {1972})}\BibitemShut {NoStop}%
\bibitem [{\citenamefont {Pauli}(1932)}]{Pauli1932HPA}%
  \BibitemOpen
  \bibfield  {author} {\bibinfo {author} {\bibfnamefont {W.}~\bibnamefont
  {Pauli}},\ }\bibfield  {title} {\bibinfo {title} {Diracs wellengleichung des
  elektrons und geometrische optik},\ }\href@noop {} {\bibfield  {journal}
  {\bibinfo  {journal} {Helv. Phys. Acta}\ }\textbf {\bibinfo {volume} {5}},\
  \bibinfo {pages} {179} (\bibinfo {year} {1932})}\BibitemShut {NoStop}%
\bibitem [{\citenamefont {Rubinow}\ and\ \citenamefont
  {Keller}(1963)}]{Rubinow1963PRev}%
  \BibitemOpen
  \bibfield  {author} {\bibinfo {author} {\bibfnamefont {S.~I.}\ \bibnamefont
  {Rubinow}}\ and\ \bibinfo {author} {\bibfnamefont {J.~B.}\ \bibnamefont
  {Keller}},\ }\bibfield  {title} {\bibinfo {title} {Asymptotic solution of the
  {Dirac} equation},\ }\href {https://doi.org/10.1103/PhysRev.131.2789}
  {\bibfield  {journal} {\bibinfo  {journal} {Phys. Rev.}\ }\textbf {\bibinfo
  {volume} {131}},\ \bibinfo {pages} {2789} (\bibinfo {year}
  {1963})}\BibitemShut {NoStop}%
\bibitem [{\citenamefont {Mur}\ and\ \citenamefont
  {Popov}(1978)}]{Popov1978SNP}%
  \BibitemOpen
  \bibfield  {author} {\bibinfo {author} {\bibfnamefont {V.~D.}\ \bibnamefont
  {Mur}}\ and\ \bibinfo {author} {\bibfnamefont {V.~S.}\ \bibnamefont
  {Popov}},\ }\bibfield  {title} {\bibinfo {title} {Semiclassical approximation
  for the {Dirac} equation in strong fields},\ }\href@noop {} {\bibfield
  {journal} {\bibinfo  {journal} {Yad. Fiz.}\ }\textbf {\bibinfo {volume}
  {28}},\ \bibinfo {pages} {837} (\bibinfo {year} {1978})}\BibitemShut
  {NoStop}%
\bibitem [{\citenamefont {Lazur}\ \emph {et~al.}(2005)\citenamefont {Lazur},
  \citenamefont {Reity},\ and\ \citenamefont {Rubish}}]{Lazur2005TMP}%
  \BibitemOpen
  \bibfield  {author} {\bibinfo {author} {\bibfnamefont {V.~Y.}\ \bibnamefont
  {Lazur}}, \bibinfo {author} {\bibfnamefont {O.~K.}\ \bibnamefont {Reity}},\
  and\ \bibinfo {author} {\bibfnamefont {V.~V.}\ \bibnamefont {Rubish}},\
  }\bibfield  {title} {\bibinfo {title} {{WKB} method for the {Dirac} equation
  with a scalar-vector coupling},\ }\href
  {https://doi.org/https://doi.org/10.1007/s11232-005-0090-1} {\bibfield
  {journal} {\bibinfo  {journal} {Theor. Math. Phys.}\ }\textbf {\bibinfo
  {volume} {143}},\ \bibinfo {pages} {559} (\bibinfo {year}
  {2005})}\BibitemShut {NoStop}%
\bibitem [{\citenamefont {Gusynin}\ and\ \citenamefont
  {Sharapov}(2006)}]{Gusynin2006PRB}%
  \BibitemOpen
  \bibfield  {author} {\bibinfo {author} {\bibfnamefont {V.~P.}\ \bibnamefont
  {Gusynin}}\ and\ \bibinfo {author} {\bibfnamefont {S.~G.}\ \bibnamefont
  {Sharapov}},\ }\bibfield  {title} {\bibinfo {title} {Transport of {Dirac}
  quasiparticles in graphene: {Hall} and optical conductivities},\ }\href
  {https://doi.org/10.1103/PhysRevB.73.245411} {\bibfield  {journal} {\bibinfo
  {journal} {Phys. Rev. B}\ }\textbf {\bibinfo {volume} {73}},\ \bibinfo
  {pages} {245411} (\bibinfo {year} {2006})}\BibitemShut {NoStop}%
\bibitem [{\citenamefont {Groth}\ \emph {et~al.}(2014)\citenamefont {Groth},
  \citenamefont {Wimmer}, \citenamefont {Akhmerov},\ and\ \citenamefont
  {Waintal}}]{kwant}%
  \BibitemOpen
  \bibfield  {author} {\bibinfo {author} {\bibfnamefont {C.~W.}\ \bibnamefont
  {Groth}}, \bibinfo {author} {\bibfnamefont {M.}~\bibnamefont {Wimmer}},
  \bibinfo {author} {\bibfnamefont {A.~R.}\ \bibnamefont {Akhmerov}},\ and\
  \bibinfo {author} {\bibfnamefont {X.}~\bibnamefont {Waintal}},\ }\bibfield
  {title} {\bibinfo {title} {Kwant: a software package for quantum transport},\
  }\href@noop {} {\bibfield  {journal} {\bibinfo  {journal} {New J. Phys.}\
  }\textbf {\bibinfo {volume} {16}},\ \bibinfo {pages} {063065} (\bibinfo
  {year} {2014})}\BibitemShut {NoStop}%
\bibitem [{\citenamefont {Nielsen}\ and\ \citenamefont
  {Ninomiya}(1981)}]{Nielsen1981NP}%
  \BibitemOpen
  \bibfield  {author} {\bibinfo {author} {\bibfnamefont {H.}~\bibnamefont
  {Nielsen}}\ and\ \bibinfo {author} {\bibfnamefont {M.}~\bibnamefont
  {Ninomiya}},\ }\bibfield  {title} {\bibinfo {title} {Absence of neutrinos on
  a lattice: (i). {Proof} by homotopy theory},\ }\href
  {https://doi.org/https://doi.org/10.1016/0550-3213(81)90361-8} {\bibfield
  {journal} {\bibinfo  {journal} {Nuclear Physics B}\ }\textbf {\bibinfo
  {volume} {185}},\ \bibinfo {pages} {20} (\bibinfo {year} {1981})}\BibitemShut
  {NoStop}%
\bibitem [{\citenamefont {Littlejohn}\ and\ \citenamefont
  {Flynn}(1991)}]{Littlejohn1991PRA}%
  \BibitemOpen
  \bibfield  {author} {\bibinfo {author} {\bibfnamefont {R.~G.}\ \bibnamefont
  {Littlejohn}}\ and\ \bibinfo {author} {\bibfnamefont {W.~G.}\ \bibnamefont
  {Flynn}},\ }\bibfield  {title} {\bibinfo {title} {Geometric phases in the
  asymptotic theory of coupled wave equations},\ }\href
  {https://doi.org/10.1103/PhysRevA.44.5239} {\bibfield  {journal} {\bibinfo
  {journal} {Phys. Rev. A}\ }\textbf {\bibinfo {volume} {44}},\ \bibinfo
  {pages} {5239} (\bibinfo {year} {1991})}\BibitemShut {NoStop}%
\bibitem [{\citenamefont {Gorbachev}\ \emph {et~al.}(2014)\citenamefont
  {Gorbachev}, \citenamefont {Song}, \citenamefont {Yu}, \citenamefont
  {Kretinin}, \citenamefont {Withers}, \citenamefont {Cao}, \citenamefont
  {Mishchenko}, \citenamefont {Grigorieva}, \citenamefont {Novoselov},
  \citenamefont {Levitov},\ and\ \citenamefont {Geim}}]{Gorbachev2014Science}%
  \BibitemOpen
  \bibfield  {author} {\bibinfo {author} {\bibfnamefont {R.~V.}\ \bibnamefont
  {Gorbachev}}, \bibinfo {author} {\bibfnamefont {J.~C.~W.}\ \bibnamefont
  {Song}}, \bibinfo {author} {\bibfnamefont {G.~L.}\ \bibnamefont {Yu}},
  \bibinfo {author} {\bibfnamefont {A.~V.}\ \bibnamefont {Kretinin}}, \bibinfo
  {author} {\bibfnamefont {F.}~\bibnamefont {Withers}}, \bibinfo {author}
  {\bibfnamefont {Y.}~\bibnamefont {Cao}}, \bibinfo {author} {\bibfnamefont
  {A.}~\bibnamefont {Mishchenko}}, \bibinfo {author} {\bibfnamefont {I.~V.}\
  \bibnamefont {Grigorieva}}, \bibinfo {author} {\bibfnamefont {K.~S.}\
  \bibnamefont {Novoselov}}, \bibinfo {author} {\bibfnamefont {L.~S.}\
  \bibnamefont {Levitov}},\ and\ \bibinfo {author} {\bibfnamefont {A.~K.}\
  \bibnamefont {Geim}},\ }\bibfield  {title} {\bibinfo {title} {Detecting
  topological currents in graphene superlattices},\ }\href@noop {} {\bibfield
  {journal} {\bibinfo  {journal} {Science}\ }\textbf {\bibinfo {volume}
  {346}},\ \bibinfo {pages} {448} (\bibinfo {year} {2014})}\BibitemShut
  {NoStop}%
\bibitem [{\citenamefont {Goerbig}\ \emph {et~al.}(2009)\citenamefont
  {Goerbig}, \citenamefont {Fuchs}, \citenamefont {Montambaux},\ and\
  \citenamefont {Pi{\'e}chon}}]{Goerbig2009EPL}%
  \BibitemOpen
  \bibfield  {author} {\bibinfo {author} {\bibfnamefont {M.~O.}\ \bibnamefont
  {Goerbig}}, \bibinfo {author} {\bibfnamefont {J.-N.}\ \bibnamefont {Fuchs}},
  \bibinfo {author} {\bibfnamefont {G.}~\bibnamefont {Montambaux}},\ and\
  \bibinfo {author} {\bibfnamefont {F.}~\bibnamefont {Pi{\'e}chon}},\
  }\bibfield  {title} {\bibinfo {title} {Electric-field--induced lifting of the
  valley degeneracy in {$\alpha$-(BEDT-TTF})$_{2}{I}_3$ {Dirac-like Landau}
  levels},\ }\href@noop {} {\bibfield  {journal} {\bibinfo  {journal} {EPL}\
  }\textbf {\bibinfo {volume} {85}},\ \bibinfo {pages} {57005} (\bibinfo {year}
  {2009})}\BibitemShut {NoStop}%
\end{thebibliography}%

\end{document}